\documentclass[conference,a4paper]{IEEEtran}
\usepackage[letterpaper, left=0.7in, right=0.7in, bottom=1in, top=0.75in]{geometry}

\usepackage{multicol}
\usepackage{graphicx}
\graphicspath{{Figures/}}
\usepackage{epstopdf}
\usepackage{acronym}
\usepackage{amsmath}
\usepackage{color, colortbl}
\usepackage{subfigure}
\usepackage[english]{babel}
\usepackage{blindtext}
\usepackage{algorithm} 
\usepackage{algorithmic} 
\usepackage{multirow}
\usepackage{color}
\usepackage{subfigure}
\usepackage{balance}
\usepackage{caption}
\usepackage{amsfonts}

\begin{document}

\newacro{3GPP}{third generation partnership project}
\newacro{4G}{$4^{th}$ generation}
\newacro{5G}{$5^{th}$ generation}

\newacro{ADC}{analogue-to-digital conversion}
\newacro{AED}{accumulated euclidean distance}
\newacro{ASE}{amplified spontaneous emission}
\newacro{ASIC}{application specific integrated circuit}
\newacro{AWG}{arbitrary waveform generator}
\newacro{AWGN}{additive white Gaussian noise}
\newacro{A/D}{analog-to-digital}

\newacro{B2B}{back-to-back}
\newacro{BCF}{bandwidth compression factor}
\newacro{BCJR}{Bahl-Cocke-Jelinek-Raviv}
\newacro{BDM}{bit division multiplexing}
\newacro{BED}{block efficient detector}
\newacro{BER}{bit error rate}
\newacro{Block-SEFDM}{block-spectrally efficient frequency division multiplexing}
\newacro{BLER}{block error rate}
\newacro{BPSK}{binary phase shift keying}
\newacro{BS}{base station}
\newacro{BSS}{best solution selector}
\newacro{BU}{butterfly unit}

\newacro{CapEx}{capital expenditure}
\newacro{CA}{carrier aggregation}
\newacro{CBS}{central base station}
\newacro{CC}{component carriers}
\newacro{CCDF}{complementary cumulative distribution function}
\newacro{CCs}{component carriers}
\newacro{CD}{chromatic dispersion}
\newacro{CDF}{cumulative distribution function}
\newacro{CDI}{channel distortion information}
\newacro{CDMA}{code division multiple access}
\newacro{CI}{constructive interference}
\newacro{CIR}{carrier-to-interference power ratio}
\newacro{CMOS}{complementary metal-oxide-semiconductor}
\newacro{CNN}{convolutional neural network}
\newacro{CoMP}{coordinated multiple point}
\newacro{CO-SEFDM}{coherent optical-SEFDM}
\newacro{CP}{cyclic prefix}
\newacro{CPE}{common phase error}
\newacro{CRVD}{conventional real valued decomposition}
\newacro{CR}{cognitive radio}
\newacro{CRC}{cyclic redundancy check}
\newacro{CS}{central station}
\newacro{CSI}{channel state information}
\newacro{CSPR}{carrier to signal power ratio}
\newacro{C-RAN}{cloud-radio access networks}

\newacro{DAC}{digital-to-analogue conversion}
\newacro{DBP}{digital backward propagation}
\newacro{DC}{direct current}
\newacro{DCT}{discrete cosine transform}
\newacro{DDC}{digital down-conversion}
\newacro{DDO-OFDM}{directed detection optical-OFDM}
\newacro{DDO-OFDM}{direct detection optical-OFDM}
\newacro{DDO-SEFDM}{directed detection optical-SEFDM}
\newacro{DFB}{distributed feedback}
\newacro{DFDMA}{distributed FDMA}
\newacro{DFT}{discrete Fourier transform}
\newacro{DFrFT}{discrete fractional Fourier transform}
\newacro{DMA}{direct memory access}
\newacro{DMRS}{demodulation reference signal}
\newacro{DOFDM}{dense orthogonal frequency division multiplexing}
\newacro{DP}{dual polarization}
\newacro{DPC}{dirty paper coding}
\newacro{DSB}{double sideband}
\newacro{DSL}{digital subscriber line}
\newacro{DSP}{digital signal processors}
\newacro{DVB}{digital video broadcast}
\newacro{D/A}{digital-to-analog}

\newacro{ECC}{error correcting codes}
\newacro{ECL}{external-cavity laser}
\newacro{EDFA}{erbium doped fiber amplifier}
\newacro{EE}{energy efficiency}
\newacro{eMBB}{enhanced mobile broadband}
\newacro{eNB-IoT}{enhanced NB-IoT}
\newacro{EPA}{extended pedestrian A}
\newacro{EVM}{error vector magnitude}

\newacro{Fast-OFDM}{fast-orthogonal frequency division multiplexing}
\newacro{FBMC}{filterbank based multicarrier }
\newacro{FCE}{full channel estimation}
\newacro{FD}{fixed detector}
\newacro{FDD}{frequency division duplexing}
\newacro{FDM}{frequency division multiplexing}
\newacro{FDMA}{frequency division multiple access}
\newacro{FE}{full expansion}
\newacro{FEC}{forward error correction}
\newacro{FEXT}{far-end crosstalk}
\newacro{FF}{flip-flop}
\newacro{FFT}{fast Fourier transform}
\newacro{FIFO}{first in first out}
\newacro{F-OFDM}{filtered-orthogonal frequency division multiplexing}
\newacro{FPGA}{field programmable gate array}
\newacro{FrFT}{fractional Fourier transform}
\newacro{FSD}{fixed sphere decoding}
\newacro{FSD-MNSF}{FSD-modified-non-sort-free}
\newacro{FSD-NSF}{FSD-non-sort-free}
\newacro{FSD-SF}{FSD-sort-free}
\newacro{FSK}{frequency shift keying}
\newacro{FTN}{faster than Nyquist}
\newacro{FTTB}{fiber to the building}
\newacro{FTTC}{fiber to the cabinet}
\newacro{FTTdp}{fiber to the distribution point}
\newacro{FTTH}{fiber to the home}

\newacro{GB}{guard band}
\newacro{GFDM}{generalized frequency division multiplexing}
\newacro{GPU}{graphics processing unit}
\newacro{GSM}{global system for mobile communication}
\newacro{GUI}{graphical user interface}

\newacro{HC-MCM}{high compaction multi-carrier communication}
\newacro{HPA}{high power amplifier}

\newacro{IC}{integrated circuit}
\newacro{ICI}{inter carrier interference}
\newacro{ID}{iterative detection}
\newacro{IDCT}{inverse discrete cosine transform}
\newacro{IDFT}{inverse discrete Fourier transform}
\newacro{IDFrFT}{inverse discrete fractional Fourier transform}
\newacro{ID-FSD}{iterative detection-FSD}
\newacro{ID-SD}{ID-sphere decoding}
\newacro{IF}{intermediate frequency}
\newacro{IFFT}{inverse fast Fourier transform}
\newacro{IFrFT}{inverse fractional Fourier transform}
\newacro{IMD}{intermodulation distortion}
\newacro{IoT}{internet of things}
\newacro{IOTA}{isotropic orthogonal transform algorithm}
\newacro{IP}{intellectual property}
\newacro{ISC}{interference self cancellation}
\newacro{ISI}{inter symbol interference}

\newacro{LDPC}{low density parity check}
\newacro{LFDMA}{localized FDMA}
\newacro{LLR}{log-likelihood ratio}
\newacro{LNA}{low noise amplifier}
\newacro{LO}{local oscillator}
\newacro{LOS}{line-of-sight}
\newacro{LPWAN}{low power wide area network}
\newacro{LS}{least square}
\newacro{LTE}{long term evolution}
\newacro{LTE-Advanced}{long term evolution-advanced}
\newacro{LUT}{look-up table}

\newacro{MA}{multiple access}
\newacro{MAC}{media access control}
\newacro{MASK}{m-ary amplitude shift keying}
\newacro{MCM}{multi-carrier modulation}
\newacro{MC-CDMA}{multi-carrier code division multiple access}
\newacro{MCS}{modulation and coding scheme}
\newacro{MF}{matched filter}
\newacro{MIMO}{multiple input multiple output}
\newacro{ML}{maximum likelihood}
\newacro{MLSD}{maximum likelihood sequence detection}
\newacro{MMF}{multi-mode fiber}
\newacro{MMSE}{minimum mean squared error}
\newacro{mMTC}{massive machine-type communication}
\newacro{MNSF}{modified-non-sort-free}
\newacro{MOFDM}{masked-OFDM}
\newacro{MRVD}{modified real valued decomposition}
\newacro{MS}{mobile station}
\newacro{MSE}{mean squared error}
\newacro{MTC}{machine-type communication}
\newacro{MUSA}{multi-user shared access}
\newacro{MU-MIMO}{multi-user multiple-input multiple-output}
\newacro{MZM}{Mach-Zehnder modulator}
\newacro{M2M}{machine to machine}

\newacro{NB-IoT}{narrowband IoT}
\newacro{NEXT}{near-end crosstalk}
\newacro{NG-IoT}{next generation IoT}
\newacro{NLOS}{non-line-of-sight}
\newacro{NN}{neural network}
\newacro{NOFDM}{non-orthogonal frequency division multiplexing}
\newacro{NOMA}{non-orthogonal multiple access}
\newacro{NoFDMA}{non-orthogonal frequency division multiple access}
\newacro{NP}{non-polynomial}
\newacro{NR}{new radio}
\newacro{NSF}{non-sort-free}
\newacro{NWDM}{Nyquist wavelength division multiplexing }
\newacro{Nyquist-SEFDM}{Nyquist-spectrally efficient frequency division multiplexing}

\newacro{OBM-OFDM}{orthogonal band multiplexed OFDM}
\newacro{OF}{optical filter}
\newacro{OFDM}{orthogonal frequency division multiplexing}
\newacro{OFDMA}{orthogonal frequency division multiple access}
\newacro{OMA}{orthogonal multiple access}
\newacro{OpEx}{operating expenditure}
\newacro{OQAM}{offset-QAM}
\newacro{OSNR}{optical signal-to-noise ratio}
\newacro{OSSB}{optical single sideband}
\newacro{OTA}{over-the-air}
\newacro{Ov-FDM}{Overlapped FDM}
\newacro{O-SEFDM}{optical-spectrally efficient frequency division multiplexing}
\newacro{O-FOFDM}{optical-fast orthogonal frequency division multiplexing}
\newacro{O-OFDM}{optical-orthogonal frequency division multiplexing}

\newacro{PA}{power amplifier}
\newacro{PAPR}{peak-to-average power ratio}
\newacro{PCE}{partial channel estimation}
\newacro{PD}{photodiode}
\newacro{PDF}{probability density function}
\newacro{PDP}{power delay profile}
\newacro{PDMA}{polarisation division multiple access}
\newacro{PDM-OFDM}{polarization-division multiplexing-OFDM}
\newacro{PDM-SEFDM}{polarization-division multiplexing-SEFDM}
\newacro{PDSCH}{physical downlink shared channel}
\newacro{PE}{processing element}
\newacro{PED}{partial Euclidean distance}
\newacro{PMD}{polarization mode dispersion}
\newacro{PON}{passive optical network}
\newacro{PPM}{parts per million}
\newacro{PRB}{physical resource block}
\newacro{PSD}{power spectral density}
\newacro{PU}{primary user}
\newacro{PXI}{PCI extensions for instrumentation}
\newacro{P/S}{parallel-to-serial}

\newacro{QAM}{quadrature amplitude modulation}
\newacro{QoS}{quality of service}
\newacro{QPSK}{quadrature phase-shift keying}

\newacro{RAUs}{remote antenna units}
\newacro{RBW}{resolution bandwidth}
\newacro{RF}{radio frequency}
\newacro{RMS}{root mean square}
\newacro{RoF}{radio-over-fiber}
\newacro{ROM}{read only memory}
\newacro{RRC}{root raised cosine}
\newacro{RSC}{recursive systematic convolutional}
\newacro{RTL}{register transfer level}
\newacro{RVD}{real valued decomposition}

\newacro{ScIR}{sub-carrier to interference ratio}
\newacro{SCMA}{sparse code multiple access}
\newacro{SC-FDMA}{single carrier frequency division multiple access}
\newacro{SC-SEFDMA}{single carrier spectrally efficient frequency division multiple access}
\newacro{SD}{sphere decoding}
\newacro{SDP}{semidefinite programming}
\newacro{SDR}{software defined radio}
\newacro{SE}{spectral efficiency}
\newacro{SEFDM}{spectrally efficient frequency division multiplexing}
\newacro{SEFDMA}{spectrally efficient frequency division multiple access} 
\newacro{SF}{sort-free}
\newacro{SIC}{successive interference cancellation}
\newacro{SiGe}{silicon-germanium}
\newacro{SINR}{signal-to-interference-plus-noise ratio}
\newacro{SISO}{single-input single-output}
\newacro{SMF}{single mode fiber}
\newacro{SNR}{signal-to-noise ratio}
\newacro{SP}{shortest-path}
\newacro{SRS}{sounding reference signal}
\newacro{SSB}{single-sideband}
\newacro{SSBI}{signal-signal beat interference}
\newacro{SSMF}{standard single mode fiber}
\newacro{STBC}{space time block coding}
\newacro{STO}{symbol timing offset}
\newacro{SU}{secondary user}
\newacro{SVD}{singular value decomposition}
\newacro{SVR}{singular value reconstruction}
\newacro{S/P}{serial-to-parallel}

\newacro{TDD}{time division duplexing}
\newacro{TDMA}{time division multiple access }
\newacro{TFP}{time frequency packing}
\newacro{THP}{Tomlinson-Harashima precoding}
\newacro{TOFDM}{truncated OFDM}
\newacro{TSVD}{truncated singular value decomposition}
\newacro{TSVD-FSD}{truncated singular value decomposition-fixed sphere decoding}

\newacro{UCR}{user compression ratio}
\newacro{UFMC}{universal-filtered multi-carrier}
\newacro{URLLC}{ultra-reliable and low-latency communication}
\newacro{USRP}{universal software radio peripheral}

\newacro{VDSL}{very-high-bit-rate digital subscriber line}
\newacro{VDSL2}{very-high-bit-rate digital subscriber line 2}
\newacro{VHDL}{very high speed integrated circuit hardware description language}
\newacro{VLC}{visible light communication}
\newacro{VLSI}{very large scale integration}
\newacro{VOA}{variable optical attenuator}
\newacro{VP}{vector perturbation}
\newacro{VSSB-OFDM}{virtual single-sideband OFDM}

\newacro{WAN}{wide area network}
\newacro{WCDMA}{wideband code division multiple access}
\newacro{WDM}{wavelength division multiplexing}
\newacro{WiFi}{wireless fidelity}
\newacro{WiGig}{Wireless Gigabit Alliance}
\newacro{WiMAX}{Worldwide interoperability for Microwave Access}
\newacro{WSS}{wavelength selective switch}

\newacro{ZF}{zero forcing}
\newacro{ZP}{zero padding}


\title{Deep Learning for Over-the-Air \\Non-Orthogonal Signal Classification}
\author{\IEEEauthorblockN{Tongyang Xu and Izzat Darwazeh}
 \IEEEauthorblockA{Department of Electronic and Electrical Engineering,
University College London, London, UK\\
 Email: {tongyang.xu.11@ucl.ac.uk, i.darwazeh@ucl.ac.uk}}}

\maketitle

\begin{abstract}
Non-cooperative communications, where a receiver can automatically distinguish and classify transmitted signal formats prior to detection, are desirable for low-cost and low-latency systems. This work focuses on the deep learning enabled blind classification of multi-carrier signals covering their orthogonal and non-orthogonal varieties. We define two signal groups, in which Type-I includes signals with large feature diversity while Type-II has strong feature similarity. We evaluate time-domain and frequency-domain convolutional neural network (CNN) models in simulation with wireless channel/hardware impairments. Simulation results reveal that the time-domain neural network training is more efficient than its frequency-domain counterpart in terms of classification accuracy and computational complexity. In addition, the time-domain CNN models can classify Type-I signals with high accuracy but reduced performance in Type-II signals because of their high signal feature similarity. Experimental systems are designed and tested, using software defined radio (SDR) devices, operated for different signal formats to form full wireless communication links with line-of-sight and non-line-of-sight scenarios. Testing, using four different time-domain CNN models, showed the pre-trained CNN models to have limited efficiency and utility due to the mismatch between the analytical/simulation and practical/real-world environments. Transfer learning, which is an approach to fine-tune learnt signal features, is applied based on measured over-the-air time-domain signal samples. Experimental results indicate that transfer learning based CNN can efficiently distinguish different signal formats in both line-of-sight and non-line-of-sight scenarios with great accuracy improvement relative to the non-transfer-learning approaches. 
\end{abstract}


\section{Introduction} \label{sec:introduction}

In legacy systems, to facilitate successful communications, both transmitter and receiver should cooperatively work on the basis of mutually-known protocols. This is at the cost of extra control overhead, time delay and inaccuracy due to the time-variant wireless channels. Therefore, non-cooperative communications are preferred in low-power low-latency communication scenarios, where a receiver can automatically distinguish signal formats.

\begin{figure*}[ht]
\begin{center}
\includegraphics[scale=0.39]{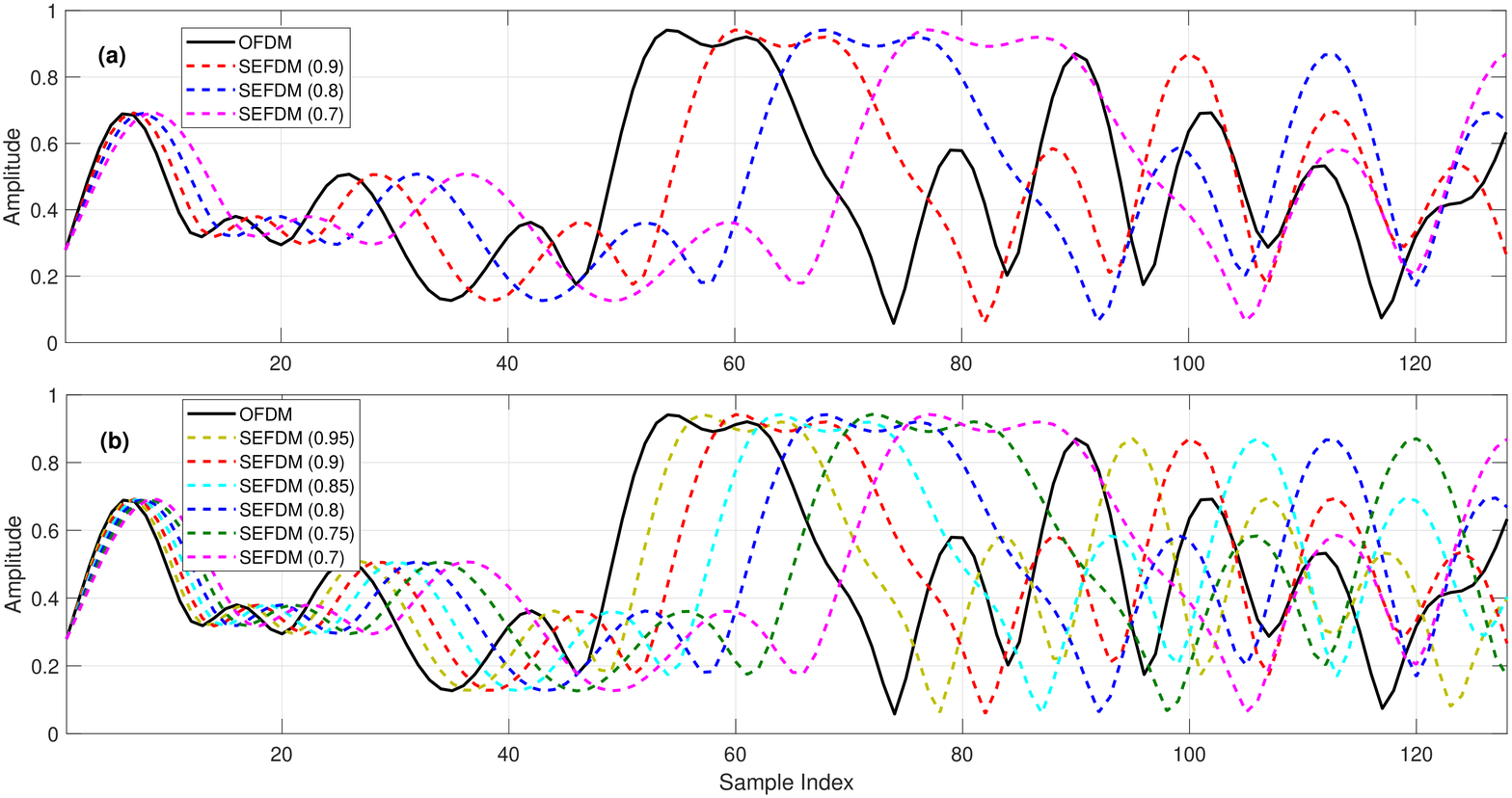}
\end{center}
\caption{Signal feature diversity and similarity visualization by modulating the same QPSK data. (a) Type-I signals. (b) Type-II signals.}
\label{Fig:CNN_feature_combined_G_I_G_II}
\end{figure*}

Recent pioneering work in \cite{OShea_classification_2018, classification_cons_TNNLS2019, AMC_TVT2018} considered the use of deep learning to extract signal features and practically revealed the possibility of using \ac{CNN} for single-carrier modulation classification. This motivated other research teams to investigate similar techniques for multi-carrier signals. Research on orthogonal signals such as \ac{OFDM} \cite{OFDM1971} has been conducted in \cite{OFDM_classification_Zhou2019ARM}, where the work showed successful classification of OFDM and single-carrier signals. More recently in \cite{OFDM_classification_access2019}, the classification of different modulated OFDM signals is explored. Due to the orthogonal sub-carrier packing feature, OFDM signals avoid internal signal interference leading to robust and accurate classification. However, for non-orthogonal signals such as frequency-domain \ac{SEFDM} \cite{Izzat_CSNDSP2018} and time-domain \ac{FTN} \cite{Anderson2013}, sub-carriers or time samples are packed closer and non-orthogonally resulting in self-created interference. This intrinsic signal interference causes ambiguity and would significantly affect signal classification accuracy. This work will focus on the spectrally efficient \ac{SEFDM}, since its flexible sub-carrier packing strategy makes it well suited for IoT communications \cite{Tongyang_NB_IoT_2018, Tongyang_MU_MIMO_NB_IoT_2018}, where non-cooperative communications will be advantageous.


Conventional \ac{CNN} models are initially trained in this work using emulation data and later are tested on over-the-air data in practical \ac{SDR} devices. {There are two reasons for the use of emulation data on the neural network training. Firstly, the training aims to extract non-orthogonal signal features, which are the common knowledge for emulation data and over-the-air data. However, obtaining emulation data is more efficient than collecting a large amount of over-the-air data. Secondly, emulation data can enhance data diversity by aggregating fast-changing channel models while real-world channels might change slowly resulting in undiversified over-the-air data. Therefore, emulation data would greatly improve the training efficiency.} The trained \ac{CNN} models work well in simulation but this might not be true for practical applications, since the training data and the real world data would have different environment features. The data feature difference is more significant in power and complexity constrained IoT communications where low cost IoT devices would be used leading to variable and non-ideal performance of transceivers hardware. Furthermore, wireless radio signals have random time, frequency and phase drifts, which would further diversify data features and complicate the neural network modelling.

Transfer learning \cite{transfer_learning_2010, transfer_learning_NIPS2014} is an approach to speed up training via fine-tuning pre-trained models. Instead of making tremendous efforts on training a single neural network to deal with multi-task problems, transfer learning extracts learnt knowledge from a source task and then applies it to a target task with fast fine-tuning according to the target task environment. This strategy is fit for precision signal classification in condition-variant over-the-air signal communications.

This work will firstly study the features of non-orthogonal multi-carrier SEFDM signals. Then eight \ac{CNN} models are trained off-line with the extensive considerations of analytical channel/hardware impairments. Moreover, an environment dependent transfer learning strategy is applied to the pre-trained \ac{CNN} models. Finally, over-the-air signal transmissions in both \ac{LOS} and \ac{NLOS} scenarios are conducted with signal classifications using the trained \ac{CNN} models and the transfer learning strategy.

The main contributions of this work are as the following.
\begin{itemize}
\item{ First time study on non-orthogonal multi-carrier signals classification using deep learning.  } 
\item{ Extensive investigations on non-orthogonal signal diversity and similarity.  } 
\item{ Over-the-air non-orthogonal signals classification.  } 
\end{itemize}

\section{Features in Non-Orthogonal Signals}

In general, signal features can be represented in time-domain samples or frequency-domain spectrum. The target SEFDM signal, in the frequency-domain, has compressed spectral bandwidth \cite{TongyangTVT2017} when compared with OFDM due to its non-orthogonal sub-carrier packing. The basic mathematical format of SEFDM signals is expressed as 
\begin{equation}
X_k=\frac{1}{\sqrt{N}}\sum_{n=1}^{N}s_{n}\exp\left(\frac{j2{\pi}nk\alpha}N\right),\label{eq:SEFDM_discrete_signal}\end{equation}
where $\alpha=\Delta{f}\cdotp{T}$ is the bandwidth compression factor, which determines the sub-carrier packing characteristics. The system is OFDM when $\alpha=1$ while $\alpha<1$ indicates SEFDM signals. Parameters $N, n, k$ are the number of sub-carriers, sub-carrier index and time sample index, respectively.

{To remove the parameter $\alpha$ in \eqref{eq:SEFDM_discrete_signal}, a new parameter $M=N/\alpha$ is introduced. By padding $M-N$ zeros at the end of each input vector (i.e. a vector consists of $N$ QPSK symbols), a new vector of input symbols is obtained as
\begin{equation}
s^{'}_{i} = \left\{
  \begin{array}{l l}
    s_{i} & \quad \text{$0{\leq}i<N$}\\
    0 & \quad \text{$N{\leq}i<M$}
  \end{array} \right.
\label{eq:single_IFFT},\end{equation} 
where the value of $N/\alpha$ has to be an integer and simultaneously a power of two, $N/\alpha\in{2^{(\mathbb{N}_{>0})}}$, which allows the IDFT to be implemented by the computationally efficient radix-2 IFFT. The SEFDM signal in a new format is expressed as
\begin{equation}
X^{'}[k]=\frac{1}{\sqrt{M}}\sum_{n=0}^{M-1}s^{'}_{n}\exp\left(\frac{j2{\pi}nk}M\right),\label{eq:FDM_discrete_signal_single_IFFT}\end{equation}
where $n,k=[0,1,...,M-1]$. The output is cut with only $N$ samples reserved and the rest $M-N$ samples are discarded. Due to the discard of the last $M-N$ samples, an SEFDM signal is essentially a partial time-domain signal representation of its OFDM counterpart.}

The time-domain samples for OFDM and SEFDM of variable bandwidth compression factors are illustrated in Fig. 1 where two types of signals are defined in the following. The number in the bracket of each item indicates bandwidth compression factors.
\begin{itemize}
\item{$\mathbf{Type-I}$: OFDM-QPSK, SEFDM-QPSK(0.9), SEFDM-QPSK(0.8), SEFDM-QPSK(0.7)} 
\item{$\mathbf{Type-II}$: OFDM-QPSK, SEFDM-QPSK(0.95), SEFDM-QPSK(0.9), SEFDM-QPSK(0.85), SEFDM-QPSK(0.8), SEFDM-QPSK(0.75), SEFDM-QPSK(0.7)} 
\end{itemize}

The same QPSK data is modulated on all the waveforms in Fig. 1 merely for signal feature diversity and similarity visualization. For realistic training and testing in the following sections, we would use random QPSK data for each signal waveform. Fig. 1(a) shows clearly the feature diversity among different SEFDM signals but with increasing similarity when signals have closer bandwidth compression factors in Fig. 1(b). Thus, classification of the Type-II signals is more challenging. The aim of this work is to automatically extract signal hidden features using \ac{CNN}. Therefore, manual feature extractions are not taken into account in this work.

Transmitted and received digital communication signals are best described as time-domain samples. Analyzing frequency-domain spectral data, additional signal processing has to be conducted for domain conversion, which is not preferred for low-power and low-latency operations. {To extensively study the diversity of performance and computational complexity, both the time-domain samples and the frequency-domain responses are investigated in this work.}

\section{Neural Network Modelling}

This work focuses on indoor communication scenarios especially for IoT applications, which have simple and relatively stable channel conditions after IoT devices deployment, but with different channels for devices  at different locations. In addition, indoor people movement would cause minor Doppler spread effect. All the impairments will be considered in the \ac{NN} modelling.

\subsection{Dataset Generation}

Work in \cite{OShea_rml_datasets_2016} provides RadioML dataset, which aims at single-carrier modulation classifications. However, for multi-carrier SEFDM and OFDM signals, new datasets have to be generated. In this work, to make neural network modelling convincing, we generate random SEFDM/OFDM samples for both training dataset and testing dataset according to the parameters in Table \ref{tab:table_signal_specifications}. {Since multi-carrier IoT signals \cite{NB-IoT-5G} prefer low order modulation formats for simplicity reasons, this work therefore focuses on QPSK modulation symbols.}

\begin{table}[ht]
\caption{Signal specifications}
\centering
\begin{tabular}{ll}
\hline \hline
$\mathbf{Parameter}$ & $\mathbf{Signal}$  \\[0.5pt] \hline 
Sampling frequency (kHz) & 200 \\ 
IFFT sample length & 2048 \\ 
Oversampling factor & 8  \\
No. of data sub-carriers & 256 \\ 
Bandwidth compression factor $\alpha$ & 1,0.95,0.9,0.85,0.8,0.75,0.7\\ 
Modulation scheme & QPSK  \\ \hline \hline
\label{tab:table_signal_specifications}
\end{tabular}
\end{table}

\begin{table}[ht]
\caption{Channel/hardware specifications}
\centering
\begin{tabular}{ll}
\hline \hline
$\mathbf{Parameter}$ & $\mathbf{Channel/Hardware}$  \\[0.5pt] \hline 
RF center frequency (MHz) & 900  \\ 
Simulation Es/N0 range (dB) & -20 $\sim$ +50  \\
Path delay (s) &  [0 9e-6 1.7e-5] \\
Path relative power (dB) &  [0 -2 -10] \\
Maximum Doppler frequency (Hz) & 4 \\
K-factor &  4 \\ 
Frequency offset (PPM) & 2 \\
Omni-directional antenna gain (dBi) & 2 \\\hline \hline
\label{tab:table_channel_specifications}
\end{tabular}
\end{table}

We emulate the analytical channel/hardware model in Table \ref{tab:table_channel_specifications} partially following the work of \cite{OShea_classification_2018, OShea_rml_datasets_2016}, in which an indoor wireless channel \ac{PDP} is defined. However, in our experiment, considering realistic indoor office environment, a time-variant wireless channel is configured with a greater maximum Doppler frequency of 4 Hz. In terms of hardware, this work uses the low-cost Analog Devices \ac{SDR} PLUTO \cite{PlutoSDR}, which is supported by Matlab. Therefore, hardware related impairments have to be reconfigured based on the PLUTO devices. According to \cite{book_PLUTO_SDR_2018}, a calibrated oscillator has a frequency offset of 2 \ac{PPM}, which will be emulated in the off-line neural network training.

\subsection{Convolutional Neural Network}

\begin{figure}[ht]
\begin{center}
\includegraphics[scale=0.32]{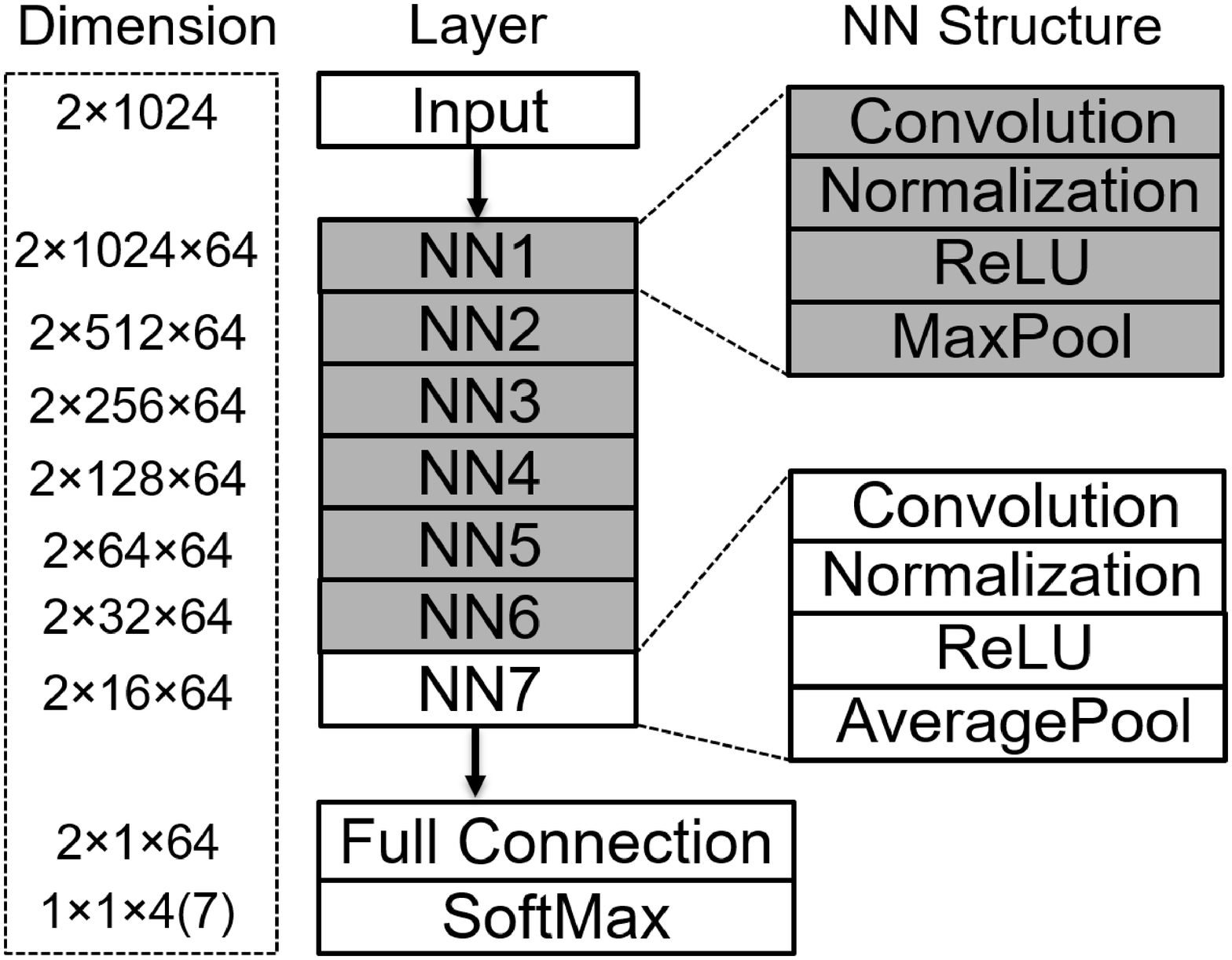}
\end{center}
\caption{CNN classifier neural network layer architecture. }
\label{Fig:CNN_architecture_classification}
\end{figure}

In this section, we train convolutional neural network models specifically for non-orthogonal multi-carrier signals classification. The accuracy of classifying testing data is the selection criterion of the optimal neural network model. For the purpose of results reproducibility, the trained \ac{CNN} layer architecture is presented in Fig. \ref{Fig:CNN_architecture_classification}, in which seven \ac{NN} layers are stacked for feature extractions. Each of the first six NN layers is made up of four sub-layers, which are presented in the grey NN structure module. In the last NN layer, the MaxPool layer is replaced by the AveragePool layer. The classification is realized by a full connection layer and a SoftMax layer with cross-entropy loss function update. The dimension of each layer is presented in the left-side column block. { It should be noted that all the following simulation and experiment results are obtained based on the CNN neural network architecture in Fig. \ref{Fig:CNN_architecture_classification}.} 

To avoid overfitting in the neural network training, a 50\% dropout ratio is set. The maximum number of epochs is limited to 30 and the mini-batch size is set to 128. To learn comprehensively from the dataset, a learning rate of 0.01 is configured.

\begin{table}[ht]
\caption{Training/validation datasets for time-domain CNN models.}
\centering
\begin{tabular}{cc}
\hline \hline
Model & Training/validation datasets    \\ \hline 
time-CNN-1  &  Type-I    \\ 
time-CNN-2  &  Type-I, channel/hardware model, Es/N0= 20 dB   \\ 
time-CNN-3  &  Type-II    \\ 
time-CNN-4  &  Type-II, channel/hardware model, Es/N0= 20 dB \\  \hline \hline
\label{tab:table_Features_training_data}
\end{tabular}
\end{table}

\begin{table}[ht]
\caption{Training/validation datasets for frequency-domain CNN models.}
\centering
\begin{tabular}{cc}
\hline \hline
Model & Training/validation datasets    \\ \hline 
fre-CNN-1  &  Fourier transform (time-CNN-1 datasets)    \\ 
fre-CNN-2  &  Fourier transform (time-CNN-2 datasets)   \\ 
fre-CNN-3  &  Fourier transform (time-CNN-3 datasets)    \\ 
fre-CNN-4  &  Fourier transform (time-CNN-4 datasets) \\  \hline \hline
\label{tab:table_Features_fre_training_data}
\end{tabular}
\end{table}

In the beginning, we designed four time-domain training/validation datasets for Type-I and Type-II signals as presented in Table \ref{tab:table_Features_training_data}. The datasets for time-CNN-1 and time-CNN-3 only consider signal intrinsic features while the other two datasets for time-CNN-2 and time-CNN-4 consider signal features, analytical channel/hardware impairments and \ac{AWGN}. In addition to the direct time-domain samples training, we also investigate the frequency-domain responses after \ac{FFT}. The frequency-domain training/validation datasets are presented in Table \ref{tab:table_Features_fre_training_data}. 

The additional computational complexity of the frequency-domain neural network classification over the time-domain neural network classification is merely the frequency response conversion FFT. The original time-domain signal is generated based on Table \ref{tab:table_signal_specifications}. To separate SEFDM/OFDM symbols from time samples and QPSK symbols, a concept of frame is used here. In this case, one frame indicates one SEFDM/OFDM symbol. Each frame has 2048 time samples with the oversampling factor of eight. At the receiver, there is no synchronization operation. The receiver would receive frames with a random time delay relative to ideal frame reception. The receiver would partially truncate 1024 consecutive time samples out of the 2048 samples. The training would operate on the truncated 1024 samples and thus without synchronization requirement. In terms of validation and testing, the receiver would also truncate 1024 samples out of 2048 samples. For the frequency-domain neural network training, the truncated 1024 time samples would firstly go through the FFT operation and then are fed to the neural network for frequency-domain training. Therefore, the additional computational complexity depends on the FFT algorithm, which is related to the length of input time samples. The number of multiplication operations of FFT is $(N_{t}/2){\cdotp}log_2{N_{t}}$ and the number of additions of FFT is $N_{t}{\cdotp}log_2{N_{t}}$, where $N_{t}$ is the number of input time samples. Since the length of input time samples is $N_{t}$=1024, the frequency-domain neural network training, per frame, requires additional 5120 multiplications and additional 10240 additions when compared with the time-domain neural network training. For validation and testing, the same number of additional multiplications and additional additions are required for each frame.

\begin{table}[ht]
\caption{Testing datasets for time- frequency-domain CNN models.}
\centering
\begin{tabular}{cc}
\hline \hline
Model & Testing datasets    \\ \hline 
time-CNN-1,2  &  Type-I, channel/hardware model,     \\
              &  Es/N0= -20:50 dB                 \\
fre-CNN-1,2   &  Fourier transform (time-CNN-1,2 datasets)   \\ 
time-CNN-3,4  &  Type-II, channel/hardware model,     \\
              &  Es/N0= -20:50 dB                 \\
fre-CNN-3,4   &  Fourier transform (time-CNN-3,4 datasets)   \\   \hline \hline
\label{tab:table_Features_testing_data}
\end{tabular}
\end{table}

\begin{figure}[ht]
\begin{center}
\includegraphics[scale=0.31]{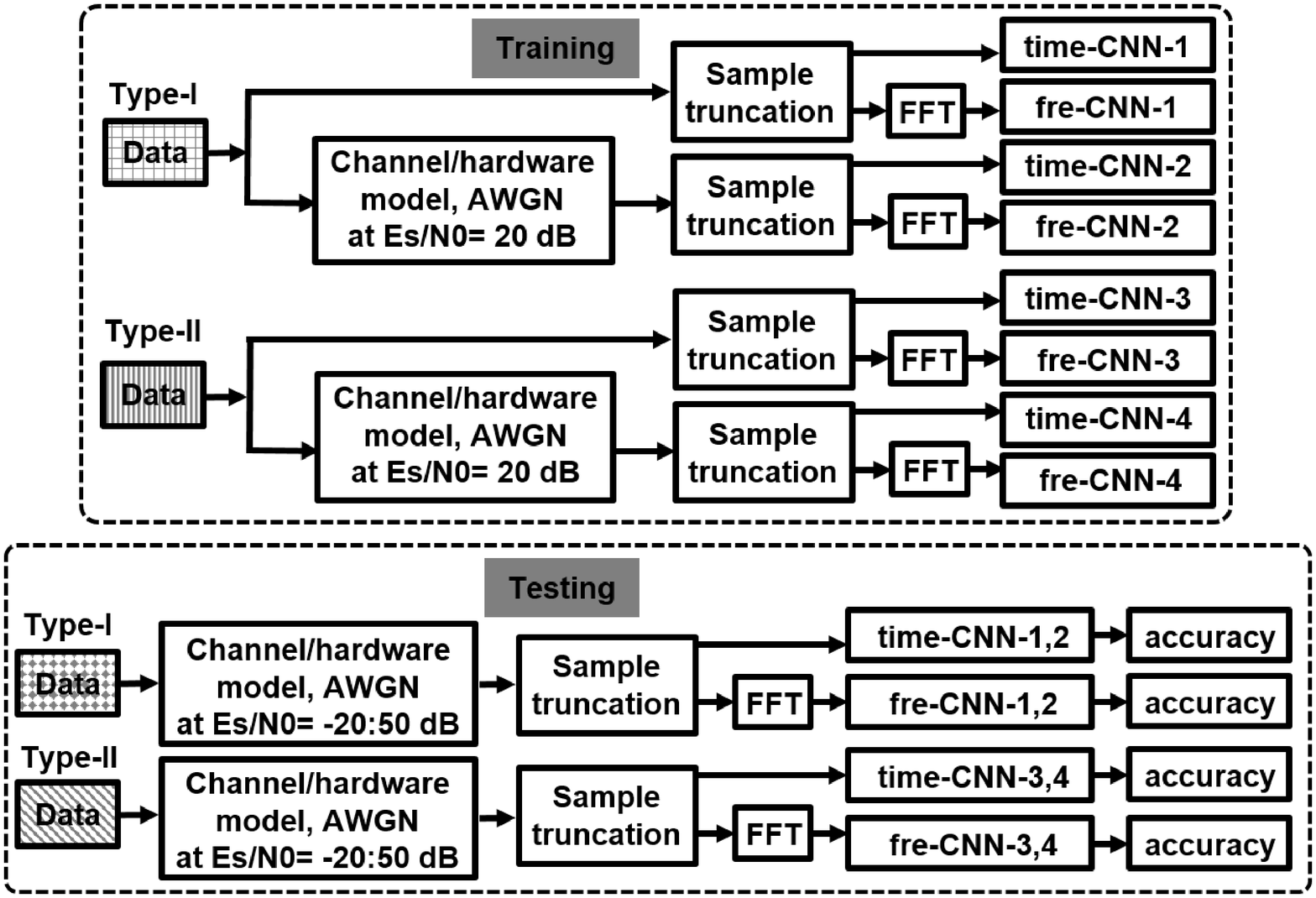}
\end{center}
\caption{Methodology of training and testing.}
\label{Fig:CNN_training_testing_methods}
\end{figure}

The block diagrams of the employed training and testing strategies are illustrated in Fig. \ref{Fig:CNN_training_testing_methods}. It clearly shows that the training is operated either with pure Type-I/Type-II data or the data with channel/hardware impairments at the fixed $Es/N0$=20 dB. However, the testing data is generated according to Table \ref{tab:table_Features_testing_data} with channel/hardware impairments and with a wide range of $Es/N0$ from -20 dB to 50 dB. Thus, the training data and testing data come from two different data sources. It should be noted that the $Es/N0$ information will not be fed to the CNN models as a training parameter. The input training information is merely the 1024 time-domain samples after the `Sample truncation' block or the 1024 frequency-domain samples after the `FFT' block.

\begin{figure}[ht]
\begin{center}
\includegraphics[scale=0.54]{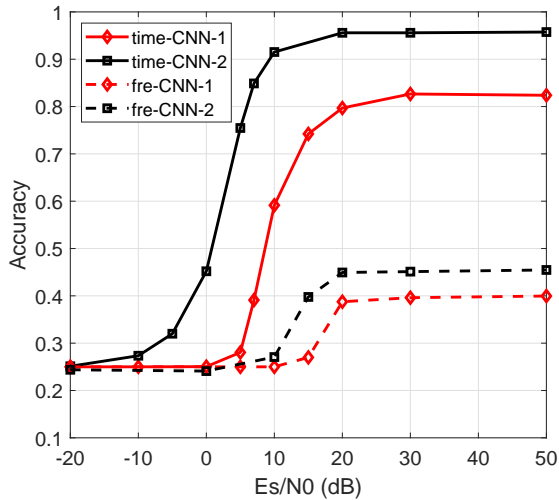}
\end{center}
\caption{Simulation signal classification accuracy for Type-I signals.}
\label{Fig:CNN_accuracy_performance_Type_I}
\end{figure}

\begin{figure}[ht]
\begin{center}
\includegraphics[scale=0.54]{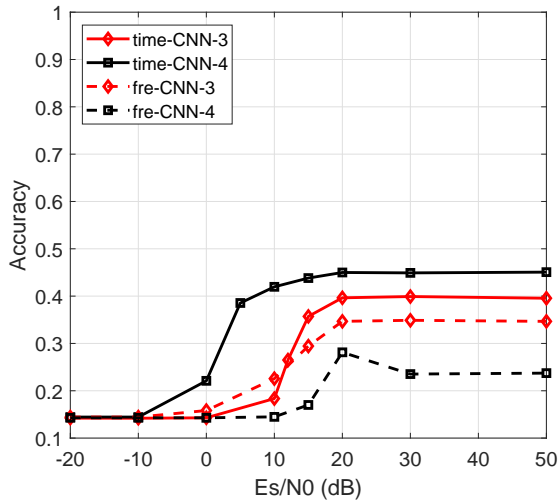}
\end{center}
\caption{Simulation signal classification accuracy for Type-II signals.}
\label{Fig:CNN_accuracy_performance_Type_II}
\end{figure}

Training is operated on an Intel(R) Xeon(R) Silver 4114 CPU (2 processors). In this work, following the information provided by Table \ref{tab:table_Features_training_data} and Table \ref{tab:table_Features_fre_training_data}, we generate 2200 frames for each signal class, in which 2000 frames are reserved for training and 200 frames are for validation. Thus, the percentages of data for training and validation are around 91\% and 9\%, respectively. In addition, a separate dataset following the information provided by Table \ref{tab:table_Features_testing_data}, consisting of 800 frames for each signal class, is used for the neural network testing. For example, for Type-I signals, there are overall 8000 frames for training, 800 frames for validation and 3200 frames for testing. For Type-II signals, there are overall 14000 frames for training, 1400 frames for validation and 5600 frames for testing. According to the training datasets defined in Table \ref{tab:table_Features_training_data} and Table \ref{tab:table_Features_fre_training_data}, four \ac{CNN} models for Type-I signals and four \ac{CNN} models for Type-II signals are trained. Testing data is independently generated according to Table \ref{tab:table_Features_testing_data}. Therefore, with the training and testing methodology defined in Fig. \ref{Fig:CNN_training_testing_methods}, the classification accuracy of the CNN models is shown in Fig. \ref{Fig:CNN_accuracy_performance_Type_I} and Fig. \ref{Fig:CNN_accuracy_performance_Type_II}.


It is clearly seen that the time-CNN-2 model, which is trained based on signals and the analytical channel/hardware model, achieves the highest accuracy. Unlike the time-CNN-2 model, time-CNN-1 is modelled using clean signals where carrier frequency offset, phase offset, time delay spread, Doppler spread, \ac{AWGN} and any other channel/hardware related impairments are ignored at the training stage. This model would be vulnerable for testing in time-variant wireless channel environments resulting in reduced accuracy as shown in Fig. \ref{Fig:CNN_accuracy_performance_Type_I}. However, for the frequency-domain CNN-1 and CNN-2 models, training and testing on frequency-domain transformed data result in significantly decreased accuracy as shown in Fig. \ref{Fig:CNN_accuracy_performance_Type_I}. It is inferred that for non-orthogonal signals, training on original time samples in deep learning CNN would gain higher accuracy than that of its frequency-domain responses.

For Type-II signals in Fig. \ref{Fig:CNN_accuracy_performance_Type_II}, due to closer bandwidth compression factors and therefore higher signal feature similarity, the accuracy levels for both time-domain and frequency-domain CNN-3 and CNN-4 are worse than those in Type-I signals. It indicates that the signal feature similarity dominates the classification accuracy in Type-II signals rather than channel/hardware condition mismatches. But the time-domain CNN models still outperform their frequency-domain counterparts.

In summary, the simulation results in this work reveal that the neural network models trained on time-domain samples lead to higher classification accuracy than the models trained on frequency-domain responses. Moreover, the frequency-domain training approach requires extra multiplication and addition operations, resulting in higher computational complexity than the time-domain training method. Therefore, the time-domain neural network training is more efficient than its frequency-domain training in both classification accuracy and computational complexity. The simulation results pave the way for the following experiment, in which only the time-domain neural network training methodology is applied.

\subsection{Transfer Learning for Signal Classification}

The accuracy of a neural network for non-orthogonal signal classification is related to {intrinsic signal features, time-frequency domain conversion and extrinsic environments.} The signal feature is deterministic once the bandwidth compression factor is fixed. However, wireless channels are time-variant in different scenarios. In addition, unexpected hardware impairments would randomly appear especially in low-cost hardware devices. This indicates that an efficient and accurate neural network model relies on either direct over-the-air data or accurate analytical channel/hardware impairments emulated data. Excessive efforts on a large amount of over-the-air data collection would be unrealistic and a single analytical channel/hardware model cannot cover all the scenarios. Therefore, the performance of signal classification is limited by model accuracy and a hence some smart learning strategy is needed.

As implied by the terminology, transfer learning \cite{transfer_learning_2010, transfer_learning_NIPS2014} {\it transfers} knowledge from pre-trained neural networks to a target task. The knowledge transfer strategy is being widely used in image and language related applications due to its faster training speed, better performance and smaller training datasets \cite{ML_handbook_2009}. In this work, we employ transfer learning in wireless communications. Work in \cite{OShea_classification_2018} follows the typical transfer learning principle \cite{transfer_learning_NIPS2014}, in which the first $d$ layers, learnt from one task network, are transferable to the first $d$ layers of another task network. The rest of network layers would be re-trained based on the target task environment. Work in \cite{DL_over_the_air_2018} studies an end-to-end deep learning system architecture, in which the transmitter, channel and receiver are aggregated together for a single neural network training. The transfer learning in this scenario would merely fine-tune the receiver side using practical over-the-air data. Another application in work \cite{transfer_learning_JSAC_2019} presents a special usage of transfer learning, in which the parameters learnt from one task network would be used for parameter initialization for the target task network. Considering the application scenario in our work, the transfer learning strategy in \cite{OShea_classification_2018} is more suitable for our applications.

Determining which part of the knowledge to be transferred plays an important role in setting the operation and accuracy of a classification function. In this work, the source task is to classify different multi-carrier signals in simulation and the target task is to classify over-the-air signals in hardware. The common knowledge (i.e part to be transferred) is the neural network architecture, which can recognize features of different signals and can be transferred to a new task. However, over-the-air data has particular channel and hardware characteristics, which include new features and have to be learnt in the new task. Since transfer learning only replaces the last few pre-trained layers, therefore the training would be faster than the initial neural network training and only a small dataset is sufficient for transfer learning. The detailed configurations of transfer learning on \ac{CNN} will be explained in the following over-the-air experiments.

\begin{figure}[ht]
\begin{center}
\includegraphics[scale=0.4]{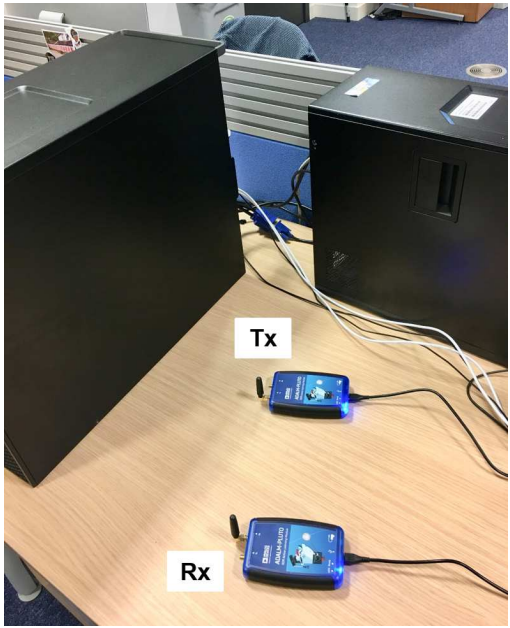}
\end{center}
\caption{Line-of-sight PLUTO experiment setup. }
\label{Fig:CNN_ML_experiments_PLUTO_LOS}
\end{figure}

\section{Experiment Design and Results}

The experiment evaluates the pre-trained time-domain CNN models on software defined radio devices PLUTO for both \ac{LOS} and \ac{NLOS} channel scenarios in an indoor office with random people movement. The PLUTO is cost-efficient; its small size makes it portable and suitable for any IoT related applications. The PLUTO SDR device is a software defined radio device mainly used for rapid idea verification. In order to make it work, Matlab software is necessary and is installed in a personal computer (PC). Therefore, the CNN training and transfer learning are both within the PC but in the SDR device. In this case, it would not cause extra power consumption of the SDR devices. Since deep learning using CNN results in complex signal processing, this work trained the CNN models off-line. Once the neural network is trained, the models would be saved. Therefore, the SDR devices would apply the saved models for online signal classification and there is no need to retrain the entire network on the device. In terms of transfer learning, only the last two neural network layers need to be retrained in Matlab on the PC to configure the SDR devices in a new channel environment. After the transfer learning, there would be no further frequent training since stable communications are assumed, in which IoT devices are stable after their initial deployments. Therefore, the off-line CNN training is a one-time operation and the transfer learning is only activated when an IoT device is re-located in a new environment. In this case, the total power consumption of the proposed scenario is reasonable when compared with traditional signal processing in wireless communications.

\subsection{Line-of-Sight Scenario}

Two PLUTO devices are placed next to each other with 30 cm distance as demonstrated in Fig. \ref{Fig:CNN_ML_experiments_PLUTO_LOS}. In addition, they are surrounded by two desktop hosts, which would introduce signal reflections. Therefore, there would be a main signal path that directly links the Tx antenna and Rx antenna with additional reflected signal paths. 

In the beginning, the pre-trained time-domain \ac{CNN} models, derived from Table \ref{tab:table_Features_training_data} are tested in the Tx-Rx communication system. We generate 800 frames for each signal class (i.e. an SEFDM signal with a specific bandwidth compression factor $\alpha$) at the Tx PLUTO device for both Type-I and Type-II signals. The second PLUTO device receives the over-the-air signal in real-time at random intervals and it truncates time samples for classification.

\begin{figure}[ht]
\begin{center}
\includegraphics[scale=0.51]{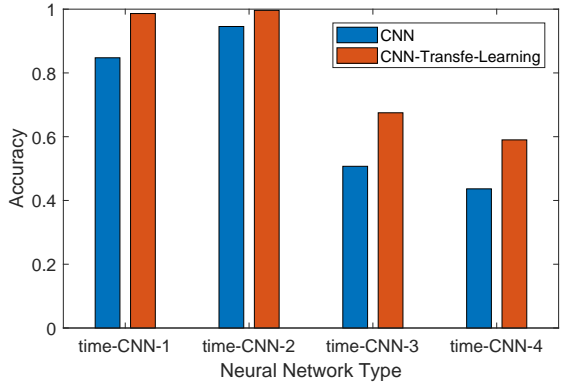}
\end{center}
\caption{Classification accuracy in the line-of-sight channel.}
\label{Fig:experiments_accuracy_results_LOS}
\end{figure}

Four pre-trained \ac{CNN} classifiers are tested and results are shown in Fig. \ref{Fig:experiments_accuracy_results_LOS}. First, similar to the results observed in Fig. \ref{Fig:CNN_accuracy_performance_Type_I} and Fig. \ref{Fig:CNN_accuracy_performance_Type_II}, Type-I signal classification has higher accuracy than that in Type-II signals. Second, in Type-I signal classification, the CNN-2 model, trained with analytical channel/hardware impairments, shows a higher accuracy level than the pure signal trained CNN-1 model of no impairments. This agrees with the simulation results obtained from Fig. \ref{Fig:CNN_accuracy_performance_Type_I}. For the Type-II signals, the pure signals trained CNN-3 outperforms the CNN-4 with impairments training. This result contradicts with the simulation results in Fig. \ref{Fig:CNN_accuracy_performance_Type_II}. It is inferred that the mutual effect of signal similarity in Type-II signals and inaccurate channel/hardware impairments modelling has greater effect on classification accuracy than for Type-I signals. This indicates that pre-processing is necessary in Type-II signals to enhance signal diversity and therefore to mitigate the mutual effect.

Considering the channel/hardware modelling mismatch between simulation and practical over-the-air radio transmissions, transfer learning is applied for fine-tuning pre-trained neural networks. Training the entire neural network is time consuming and unrealistic for practical scenarios since a wireless channel would change frequently. Therefore, in this work only the last two layers of Fig. \ref{Fig:CNN_architecture_classification}, concerned with extraction of channel features, are replaced; namely the full connection layer and SoftMax layer. Transfer learning requires new datasets input to fine-tune the pre-trained neural network. In the beginning, the receiver side PLUTO will collect 50 frames per signal class for re-training the last two layers to learn practical over-the-air channel/hardware knowledge. Since only the last two layers have to be re-trained, the entire transfer learning would be much faster. Practical results reveal that transfer learning can significantly improve classification accuracy levels for four CNN models. For Type-I signals, the CNN-1 and CNN-2 models reach almost 100\% accuracy. For Type-II signals, both CNN models are improved via the use of transfer learning by up to 35\%, but are still influenced strongly by the imperfect models of the channel and hardware, and therefore give largely similar performance.

\begin{figure}[ht]
\begin{center}
\includegraphics[scale=0.4]{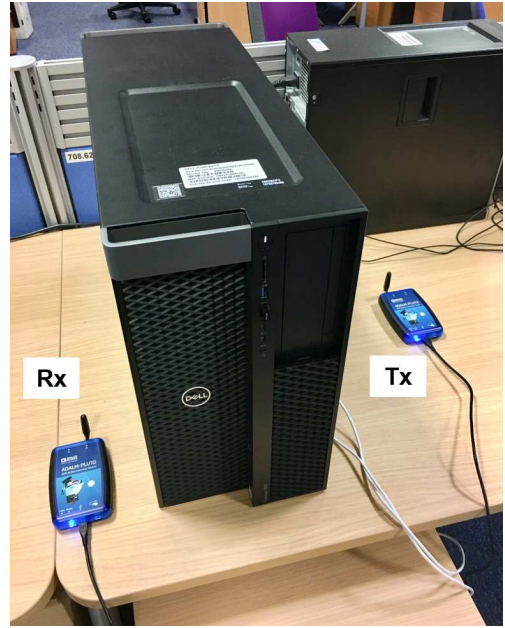}
\end{center}
\caption{Non-line-of-sight PLUTO experiment setup. }
\label{Fig:CNN_ML_experiments_PLUTO_NLOS}
\end{figure}

\begin{figure}[ht]
\begin{center}
\includegraphics[scale=0.51]{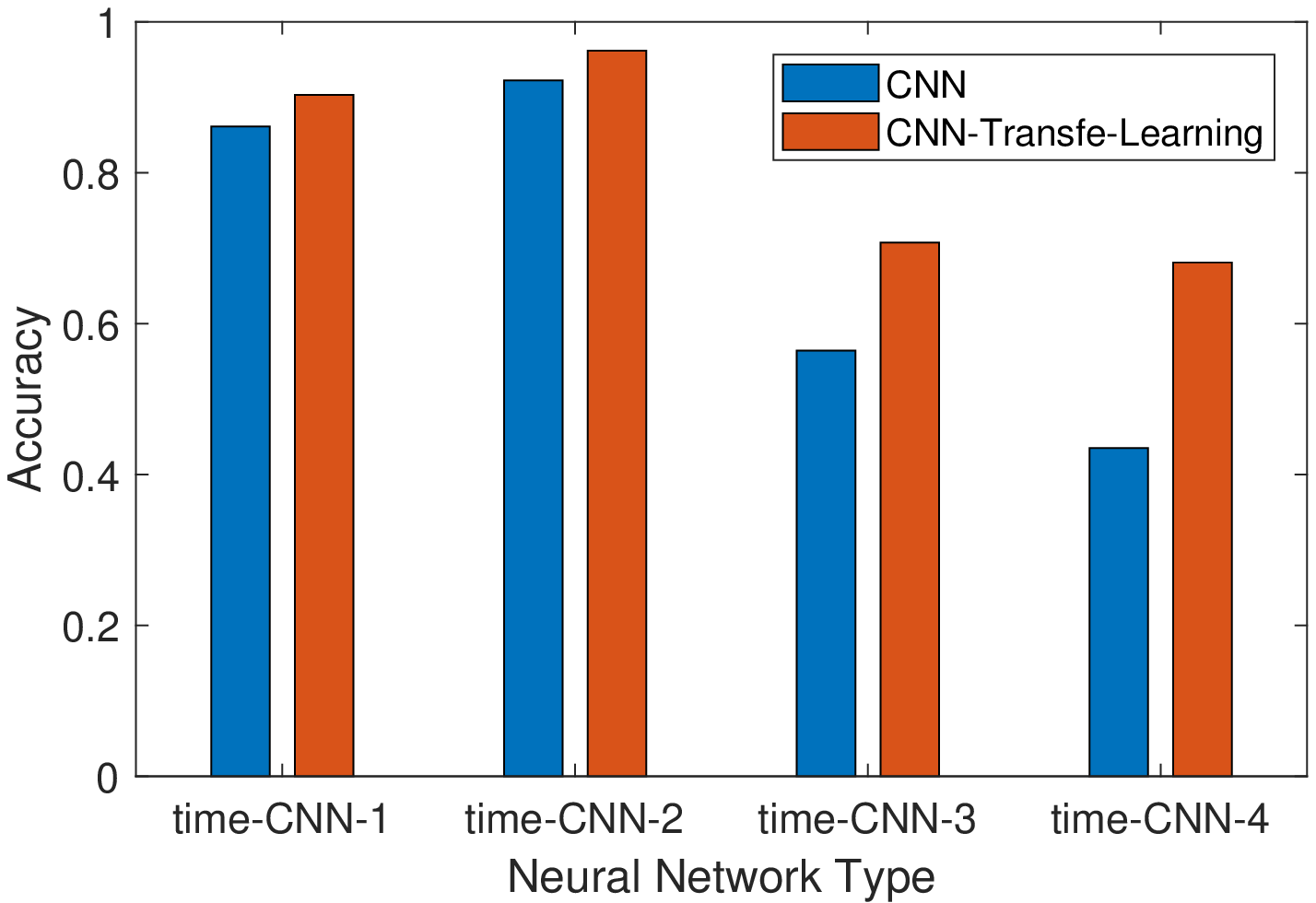}
\end{center}
\caption{Classification accuracy in the non-line-of-sight channel.}
\label{Fig:experiments_accuracy_results_NLOS}
\end{figure}

\begin{figure*}[ht]
\begin{center}
\includegraphics[scale=0.46]{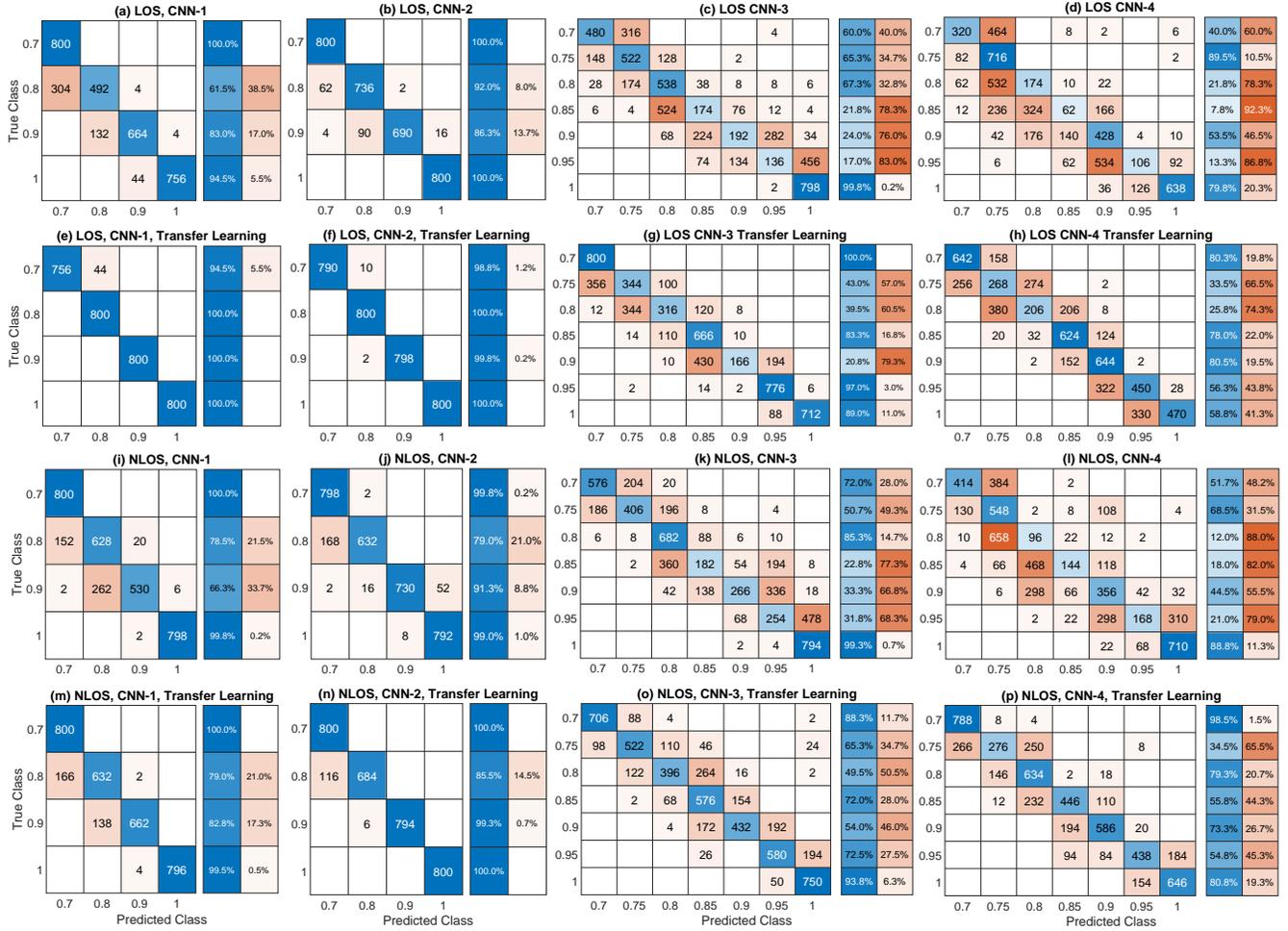}
\end{center}
\caption{Confusion matrix visualization. Type-I signal classification: (a,b,e,f,i,j,m,n). Type-II signal classification: (c,d,g,h,k,l,o,p). }
\label{Fig:full_confusion_matrix_LOS_NLOS}
\end{figure*}

\subsection{Non-Line-of-Sight Scenario}

To evaluate the robustness of the trained CNN models in a wide range of scenarios, \ac{NLOS} signal communications are set up in Fig. \ref{Fig:CNN_ML_experiments_PLUTO_NLOS} via placing obstacles between the transmitter and receiver. Results in Fig. \ref{Fig:experiments_accuracy_results_NLOS} reveal that the classification accuracy levels for Type-I signals are still higher {than those} of Type-II signals even with obstacles blocking signal propagation. Applying transfer learning, the accuracy is further improved for Type-I and Type-II signals by up to 57\%.

\subsection{Confusion Matrix Comparisons}

Table \ref{tab:table_classification_accuracy_percentage_LOS_NLOS} summarizes the numerical classification accuracy results for the different CNN models, communication scenarios and system testbeds.

To visualize the classification accuracy for each signal format, confusion matrices are illustrated in Fig. \ref{Fig:full_confusion_matrix_LOS_NLOS}, in a similar representation to that of \cite{OShea_classification_2018}. In each sub-figure, classes indicate compression factors $\alpha$, vertical labels indicate true transmitted signal classes and horizontal labels indicate predicted signal classes. Perfect signal classification would show only diagonal elements in each confusion matrix. Therefore, it is visually concluded that Type-I signals yield higher classification accuracy than Type-II signals. The reason for this has been explained in Fig. 1, in which Type-II signals have stronger signal similarity leading to false classifications.

\begin{table}[ht]
\centering
\caption{Classification accuracy for LOS and NLOS channels}
\begin{tabular}{|c|c|c|c|c|}
\hline
\multirow{2}{*}{Model}  & \multicolumn{2}{c|}{LOS} & \multicolumn{2}{c|}{NLOS}    \\   \cline{2-5} 
                        & Direct & Transfer learning     & Direct & Transfer learning   \\ \hline \hline   
CNN-1  &  84.75\%   &  98.63\%  &  86.13\%   &  90.31\%   \\
CNN-2  &  94.56\%   &  99.63\%  &  92.25\%   &  96.19\%   \\
CNN-3  &  50.71\%   &  67.50\%  &  56.43\%   &  70.75\% \\ 
CNN-4  &  43.64\%  &  59.00\%   &  43.50\%  &  68.11\% \\  \hline
\end{tabular}
\label{tab:table_classification_accuracy_percentage_LOS_NLOS}
\end{table}

It should be noted that the use of transfer learning can efficiently mitigate the channel/hardware mismatch between analytical models and practical models. However, for Type-II signals, due to the feature similarity, signals are easily classified in error into adjacent signal classes. This could be mitigated via extra signal processing prior to signal classification.

\section{Conclusion}

This work deals with an intelligent signal classification task for non-orthogonal SEFDM signals in both simulation and over-the-air experiments. Unlike interference-free single-carrier and orthogonal multi-carrier OFDM signals, the sub-carriers within SEFDM are non-orthogonally packed leading to higher spectral efficiency at the cost of self-created interference. Therefore, classifying different SEFDM signals would be more challenging, which is the aim of this work. Deep learning is applied for the classification in this work, where convolutional neural network (CNN) models, both in the time-domain and frequency-domain, are specifically designed and trained for SEFDM. Simulation results verify that the time-domain CNN models outperform their frequency-domain models in both classification accuracy and computational complexity. Further results reveal that classification accuracy is improved when a CNN is trained with data derived from a group of signals with wide variation of their compression factors (i.e signals with strong signal diversity). Using software defined radio devices, experimental work, with practical over-the-air testing, is conducted in LOS and NLOS scenarios for various CNN models pre-trained on data with and without channel and with varying levels of spectral efficiencies. Measured results verify the pre-trained models and compare performance in terms of accuracy and confusion matrices. To improve accuracy and deal with the problem of mismatch between analytical and practical wireless channel and hardware impairments, a practical transfer learning strategy is applied to fine-tune the pre-trained models, showing classification accuracy improvement up to 57\%, depending on the application scenario used. The classification accuracy of the conducted experiments, when using the specially designed transfer learning strategy, ranged from nearly 60\% to nearly 100\%. In summary, this proof of concept work has shown experimentally and by numerical simulations the efficacy of using deep learning techniques to classify non-orthogonal multi-carrier signals with varying levels of inter-carrier interference. To improve classification accuracy, further research into signal processing has to be undertaken to amplify signal diversity and to derive accurate channel/hardware impairments models for robust neural network training.


\bibliographystyle{IEEEtran}
\bibliography{Tongyang_Ref}

\end{document}